\documentclass[nolinenum]{jfp}

\def\mnd{\par\medskip\noindent} \def\snd{\par\vskip6pt\noindent}
\usepackage{url}
\def\itbf{\bfseries\itshape}
\usepackage{fancyvrb}
\usepackage[utf8]{inputenc}
\usepackage{graphicx}
\usepackage{wrapfig}

\font\cour=pcrb8r at 10pt
\def\tt{\cour}
\font\scour=pcrb8r at 9pt
\makeatletter
\def\verbatim{\small\@verbatim\frenchspacing\@vobeyspaces\@xverbatim}
\def\verbatim@font{\cour}

\makeatother
\definecolor{gry}{rgb}{0.7, 0.4, 0.4}
\def\ovl#1{{\overline#1}}

\begin{document}

\journaltitle{JFP}   
\cpr{Cambridge University Press}   
\doival{10.1017/xxxxx}   
\jnlDoiYr{}   

\title{Corecursive coding of High Computational Derivatives and Power Series}
\begin{authgrp}
 \author{Jerzy Karczmarczuk}
 \affiliation{University of Caen, France (retired, associated)}
 \today
\end{authgrp}
	
\begin{abstract}
We discuss the functional lazy techniques in generation and handling of arbitrarily long sequences of derivatives of numerical expressions in one ``variable''; the domain to which the paper belongs is usually nicknamed ``Automatic differentiation''. Two models thereof are considered, the chains of ``pure'' derivatives, and the infinite power series, similar, but algorithmically a bit different. We deal with their arithmetic/algebra, and with more convoluted procedures, such as composition and reversion. Some more specific applications of these structures are also presented.  
\snd
\end{abstract}
\maketitle

\section{Introduction}

\noindent
The essential part of this paper is the implementation and some applications of high-order derivatives within the Automatic Differentiation approach (\cite{WickAD}, see also \cite{Autodiff}, and hundreds of other sources), 
the {\em numerical} computation of derivatives inferred from the typical, standard scientific/numerical programs\footnote{For historical reasons, in several texts for beginners, the term: ``automatic'' differentiation is {\em opposed} to ``numerical'', which repeatedly means the approximations through difference quotients. We disagree with this perception of the term ``numerical''.}.  However, since there is rarely something really automatic therein, we prefer to employ the word ``computational''), using the lazy programming techniques in Haskell. We shall concentrate on the 1-dimensional domain, and the forward AD mode. It is assumed that the Reader is (at least superficially) acquainted with this domain; if needed, the references are reasonably copious. This is not a new subject, we wanted to treat some more sophisticated examples than usually offered. 

This is a development of our work (\cite{Hosc}; see also \cite{Icfp}), and although we present here some new contributions to the subject, the main purpose of this text is pedagogic: we want to show the power, and the readability of some practical corecursive programming {\em techniques}, beyond the introductory level. However, we stay within the numerical computing craft.

We adapted some constructs to more recent versions of Haskell, and we show a remarkably simple and compact coding of the derivative train (chain) composition and reversion\footnote{We present numerical algorithms, noticeably shorter than symbolics, treated in several other papers.}. We complement the subject by a discussion of the coding of infinite power series, especially in relation with the first part. This last model has also been treated elsewhere, also functionally, see e.g. the instructive papers of Douglas McIlroy, (\cite{DougP1,DougP2}), better known than ours (\cite{LPower}). Our cited paper treats some more complicated issues, but the inventivity and the pedagogical talents of Doug McIlroy should be acknowledged. BTW, several techniques of the processing of series can be found in the second volume of {\em The Art of Computer Programming} (\cite{Knuth}); the comparison of the imperative array processing by Knuth with the functional code, may be inspiring. See also the book (\cite{Schwatt}). The affinity between the concerned datatypes: series and sequences of derivatives (chains) is significant, but there are some algorithmic differences which are brought up.

Since the purpose of this writing was mainly pedagogical, the material was chosen, so that it should be fairly comprehensive, but readable; we had to include some introduction to standard (but not always taught) functional concepts, such as {\itbf Functors}, and also {\itbf Foldables}, but without details. Abstractions and generalisations should come after acquiring some familiarity with the {\em lazy algorithmisation} based on the corecursive, ``runaway'' recursion, permitting to see equations as algorithms, in a style different from the imperative one.

\mnd
In natural sciences, the first two derivatives (in time or in 
other\footnote{In relativistic physics other dimensions may be more appropriate. Also in the s.c. survival analysis, the time is often a dependent variable.}  
principal independent parameter) of dynamical variables, dominate over the higher differentiation constructs, since the differential equations: of motion, flows, or of other quantities resulting from the {\em typical} dynamics, are second order. The typical  {\em stability estimations}, unavoidable in many optimisation calculi, need as well the two first derivatives. The machine learning computations belong to this category too, and all this influenced also the teaching of the application of mathematics\ldots{} Unfortunately, a fair percentage of users of applied mathematics, and also of teachers, consider the subject of higher derivatives not too interesting\footnote{What is so fascinating in the 5-th derivative, it is just the derivative of the 4-th, etc., so you know them all\ldots}. So, although the mathematical literature on the formal issues is reasonably abundant, there are hundreds of pages devoted to the discussion of the formula of Faà di Bruno, (\cite{FaaDiBruno,Johnson}), it is not easy to find much material applied to coding. 

But, of course, we need those derivatives. Some state equations in modern physics require them (\cite{Visser}). They will appear when our system model treats the dispersion phenomena (e.g., rainbows, solitons, etc.) and radiative corrections, and this is not just a high-brow theoretical issue. The modelling, planning and analysis of the highly nonlinear beam dynamics in accelerators require high derivatives, and this is ``just'' engineering (\cite{MBerz}).
The {\em changes} of acceleration (jerks) are unpleasant for humans, and this conditions the design of roller-coasters, railways, vehicle acceleration devices, etc. The third time-derivative has even an official (English) name: ``jerk'' (\cite{Eager}, and there were others: jolt, surge and lurch\ldots). Higher entity names became rather anecdotical: snap, crackle, pop\footnote{They are names of ``gnomes'': mascots decorating the Kellogg's Rice Krispies boxes. (\cite{Crackle}). Thus, the issue is \ldots{} uhm, highly practical.}. 

On the other hand, formal mathematical studies deal with much higher (or even an indeterminate number of) derivatives. Power series development is ubiquitous.  Some techniques of solving, or improving the convergence of the differential equations, exploit them quite often. The (asymptotic) series resulting from the perturbational calculi may behave badly, and a good, sometimes weakly known number of terms may be useful e.g. for their Padéization or the convergence estimation. There are differential recurrences, e.g. the chain of Hermite function values may be computed as:
\begin{equation}
H_n(x) = \frac{1}{\sqrt{2n}} \left(x H_{n-1}(x) - \frac{d H_{n-1}(x)}{dx} \right)\,, \qquad\text{where}\qquad H_0(x)=e^{-x^2/2}\,.  \label{hermit}
\end{equation}
Historically it was uncommonly exploited, since demanded supplementary manual work. 

In mechanics (or thermodynamics) one needs to invert the Legendre transforms, where functions sometimes come from the perturbative expansions, and take the form of series, or arrays of derivatives. 
The examples are numerous; in any case, the construction and manipulation of higher derivatives is a never-ending story, the publications thereon continue until today, and it might be interesting to focalize on algorithms which abstract from the maximum order (the number of terms) we need, or suspect that we need\footnote{Until the moment where the values must be  printed or plotted, or instantiated otherwise.}. Here, the lazy streams may be of assistance, and it seems that they are still underappreciated by the community interested in numerical computations.

Our main target audience are programming folk working in applied math domain, and students. The expected Readers' level should permit them to understand short Haskell programs. We tried to avoid more esoteric abstractions, adequate for the accomplished ``functional'' computer scientists. Our programs have been tested under {\tt ghci}, on GHC version 9.8.  All code source described in this text is available from the author\footnote{Since the programs are unstable, please contact JK.}. 

The importance of the computational differentiation is known and discussed many times. We underline the fact, that although for first derivatives we have good, inexpensive, and elegant finite-difference approximations, (such as the complex step algorithms, see e.g., (\cite{Squitra,CsMartins}), such approach to high derivatives is usually ill-conditioned and awkward.

\section{Derivative chains (or towers) in Haskell}

\noindent
As said, we shall describe the derivative chains and series; this section treats the chains, and the manipulation of series is deferred to the section \ref{serpow}.
The well known technique of Clifford {\em dual numbers}, tuples $(y,d)$ extends the numerical expressions into structures, whose one member, here: $y$, is the ``main value'', and $d$ -- the derivative; the {\em differentiation variable} is implicit, it is an object $(x,1)$ bearing {\em any} name in the code; it should be unique, while all {\em constants}, manifest or derived, should be of the form $(c, 0)$. The generalisation of the arithmetic operations thereupon are straightforward, e.g., $(a,b)+(c,d)=(a+c,b+d)$, $(a,b)\cdot(c,d)=(ac,ad+bc)$, etc. The main weakness of these tuples -- from our perspective -- is that the building of higher derivatives is not {\em naturally} workable; an expression and its derivative belong to different domains.

Our intention was to enhance the computational domain to objects which may serve as elements of a closed {\em differential algebra} (\cite{Ritt}; see also \cite{Kapla}), where the derivative of an expression is an instance of the same species. 

The field of numbers is augmented by the {\em derivation} operator, which should be linear, respect the Leibniz rule, be coherent with the chain rule, etc. We follow the commonly\footnote{But not always} employed protocols: the derivation operator acts on expressions, not on functions, and it is a {\itbf structural entity}; we don't think about a {\em differentiation operator}, but rather about derivatives -- constituents of our data, which are assembled by the operations beneath the user source code\footnote{Call them {\em automatic}, but all other operations controlled by the compiler and runtime librairies are much alike.}.

So, if $e=(y,d)$, then $d$ must be $(z,f)$, and so on. The derivative is always a compound, so it cannot be (logically) finite, although it might be physically cyclic. In a nutshell, we shall work with sequences equivalent to infinite lists: $[y, y^{(1)}, y^{(2)},y^{(3)},\ldots]$. The term ``the derivative'' without further qualification, within such form denotes rather $[y^{(k)},y^{(k+1)},\ldots]$, than $y^{(k)}$. This one is the main value of the bracketed object above.

\mnd
The representation chosen for our extended numerical expressions and their arithmetic will not be based oh Haskell lists, in order to avoid errors, since we use standard lists as well.

For the chosen data structures we will employ a handcrafted {\em infinite} list-like sequences, as {\tt p :> q :> r $\cdots$}, where {\tt p} etc., are values of numerical expressions; every one is followed by its derivative. The constructor \verb|(:>)| replaces {\tt (:)}. Such chains are ``infinite'' in the sense: no ``empty'' object similar to {\tt []}, which would end the list. (Actually, we optimised a little the structure, we {\em have} finite data as well, but they are not fundamental, and they are just mentioned.)
\vspace{-0.25cm}
\begin{Verbatim}[fontfamily=courier,fontseries=b,commandchars=\\\{\}]
data Df a = C !a | !a :> Df a
dcst c = c:>zeros      \itshape\color{gry} -- A constant as a stream
dvar x = x :> dcst 1   \itshape\color{gry} -- The variable
\end{Verbatim}
\vspace{-0.25cm}
The variant {\tt C p} denotes a constant, semantically equivalent to {\tt p :> 0 :> 0 :> \ldots} This is the optimisation mentioned above, which avoids towers of zeros (which, if we are careful, are not very costly, because of laziness), but for testing it is useful.

As said, if the second element of such a sequence is equal to 1, and all remaining vanish, this is the extensionally defined {\itbf differentiation variable}. Further on, {\em in this text} we shall omit the variant C, but it should not be entirely forgotten, see the Hermite function example on page \pageref{hermdef}. 

The derivation yields the tail of the chain, but calling the access of a data part a ``structural'' action, may be a misnomer, since accessing an item of a lazy structure may imply other, perhaps costly operations. It should be noted that we {\em (almost) never compute explicitly the derivative of an expression}; we perform the arithmetic with the complete towers of values, the appropriate values are generated (co)recursively ``behind the scene'', when we demand the result.
The {\bf corecursivity} means that the recursion takes place within the function codomain, when new data chunks are {\em produced}, like in the definition of the stream known as {\tt [n ..]} in Haskell: {\tt ints n = n : ints (n+1)}, yielding $n, n+1, n+2,\ldots$

What was essential for us, was the lazy scheme: open, {\em extrapolating} recursion, without any control of loop limits, since the length of generated sequences is meaningful only during their incremental {\em consumption}: if one wants only 10 terms, it suffices {\itbf  not to access} the 11-th or further.

The addition of expressions is linear, and the multiplication obeys the Leibniz rule:
\begin{equation}
x@(x_0 :> x') \cdot y@(y_0 :> y') = x_0 y_0 :> x y' + x' y\,.  \label{trivmul}
\end{equation}
Of course, {\tt(+)}, and {\tt(*)} should be defined within the {\tt Num} instance of the chain type, but before we return to the above definition, notice that we find here such terms as $x y'$, and {\em not}:  $x_0 y'$; at the RHS of (\ref{trivmul}). We shall return to this definition, and to the division slightly later, since they {\itbf have to} be optimised; here are the definitions of the standard arithmetic, to place in the instances of such classes as {\tt Fractional} and {\tt Floating}.
\vspace{-0.25cm}
\begin{Verbatim}[fontfamily=courier,fontseries=b,commandchars=\\\{\}]
\itshape\color{gry}... Auxiliary (sample)
\color{gry}dmap f (x0:>x') = f x0 :> dmap f x'
\color{gry}c *> x = dmap (c*) x    \itshape\color{gry} -- See later...
dzip f  (x0:>x') (y0:>y') = f x0 y0 :> dzip f x' y'
dzip3 [] _ _ = 0    -- Finite, inhomog., 2 towers and a list
dzip3 (b0:bq) (p0:>pq) (q0:>qq)=(b0*p0*q0) + dzip3 bq pq qq
sqr x = x*x
sincos x@(x :> x') = (a,b) where   \itshape\color{gry} -- Optimisation
    a = sin x :> x'*b;  b = cos x :> (-x'*a)
\itshape\color{gry}... In instances
recip (x0:>x') = ip where ip = recip x0 :> (-x'*sqr ip) 
exp (x0:>x') = w where w = exp x0 :> (x' * w)
log (x@(x0:>x')) = (log x0 :> (x'/x))
sqrt (x0:>x') = w where w = sqrt x0 :> ((1/2) *> (x'/w))
sin xs=fst (sincos xs)
cos xs=snd (sincos xs) \itshape\color{gry}--If needed both, factor sincos out.
tan (x0 :> x') = w where w = (tan x0 :> x'*(1+w*w))
atan (x@(x0:>x')) = atan x0 :> (x'/(1+sqr x))
asin x@(x0:>x') = asin x0 :> x'/sqrt(1-sqr x)
acos x@(x0:>x') = acos x0 :> (-x'/sqrt(1-sqr x))
\end{Verbatim}
\vspace{-0.25cm}

\subsection{Digression: enhancing abstraction}

\noindent
This fragment may interest more advanced Readers, and may be important, although doesn't change much in {\em this} text\ldots{} 

We have chosen a new sequential, iterable data structure linked with {\tt (:>)}, since we wanted to have object similar to lists, but really different. This meant, that we had to define several auxiliary operations, which cloned the existing functionalities, e.g. {\tt dmap} which replicates the standard list functional {\tt map}, also: the ``zips'', etc. This implied also the intense usage of recursive procedures, and several of them employ the frequent, well known recursive paterns, defined already in languages which preceded Haskell.

But Haskell evolves, and its controlled polymorphic features based on type classes, become more rich, and their integration with the user layer -- more comfortable. Every student knows that we can emulate numbers, i.e., imagine and build objects which we can multiply, divide, store in graphic form, etc., and profit from the standard parsers. The overloading of arithmetic is the very foundation of our differentiation package.

Now we can overload {\itbf mappings}, and also {\itbf folds} for the user data types. If we put in our program
\vspace{-0.25cm}
\begin{Verbatim}[fontfamily=courier,fontseries=b,commandchars=\\\{\}]
\{-# LANGUAGE DeriveFoldable #-\}
...
data Df a = C !a | !a :> Df a  deriving  (Functor,Foldable)
...
c *> x = fmap (c*) x \itshape\color{gry}--This is enough, no dmap necessary.
\end{Verbatim}
\vspace{-0.25cm}
the system enables the inheritance of the overloaded ``map'' called {\tt fmap}, so no definition of ``dmap''  is necessary, and the option in the preamble allows also the derivation of {\tt foldr}, {\tt foldl} etc.

Since such useful functionals as {\tt sum} or {\tt product} apply the folding procedures, wthout anything else,{\tt product (11 :> 22 :> 3 :> C 3)} becomes executable, and yields 2178. We remind the definition of the right folding, 
\vspace{-0.25cm}
\begin{Verbatim}[fontfamily=courier,fontseries=b,commandchars=\\\{\}]
foldr op ini (x:xq) =  x `op` (foldr op ini xq)
\end{Verbatim}
\vspace{-0.25cm}
which combines all elements of a list, or some other sequential data collection, iterating a binary operator {\tt op}, beginning with an initial value {\tt ini}. The function {\tt product} is defined as {\tt product l = foldr (*) 1 l}, and, of course, cannot reduce an infinite list. We return to this subject in the section \ref{cival}.

\subsection{The  Lambert ``W'' function}

\noindent
If one knows how the derivative depends on the function itself (or perhaps on known others, as in the case of trigs and cyclometric functions), the construction of the full tower is easy, the self-referring examples above show how. This list can be extended, a meaningful example is the Lambert ``W'' function (\cite{LambertW}), given e.g., through the implicit equation
\begin{equation}
W(x) \cdot e^{W(x)} = x\,, \quad\text{which implies}\quad.W'(x) = \frac{\exp(-W(x))}{1+W(x)} = \frac{W}{x\cdot(1+W)}\,, \label{wladf}
\end{equation}
where the second form requires that the point $x=0$ be treated separately. 
If one just wants to see its plot, it suffices to take its reverse from the formula (\ref{wladf}) above, and draw it, exchanging the axes.
We shall find all derivatives of the $W$ branch regular at zero, where it must vanish. It fulfils the equation:
\vspace{-0.25cm}
\begin{Verbatim}[fontfamily=courier,fontseries=b]
lw0 = 0 :> exp(-lw0) / (1+lw0)
\end{Verbatim}
\vspace{-0.25cm}
which gives immediately the chain [1.0, -2.0, 9.0, -64.0, 625.0, -7776.0, 117649.0, -2097152.0,\ldots], i.e., $(-n)^{n-1}$, from $n=1$ up. The Reader is invited to compare this result with the formulae in the paper cited above. We shall return to ``W'' in the context of series, section \ref{lambsect}. \label{lambco} However, this is not an obscure (and marginally exotic, since rarely taught) special function; its importance in sciences is meaningful (\cite{Brito,Mezo}).

The lazy corecursive construct with self-references is not just the elegance  of the ``infinite'' stream implementation, but also the conversion of a {\em fixed-point equation}  into an effective algorithm, which is not difficult to code. This is an important pedagogic purpose of this work. We exploited these techniques, when we tried to promote functional programming among students of physics, and engineering; functional algorithmisation is more frequently adapted to numerous problems appearing in mathematically-oriented natural sciences, than those communities expect! For example, the generation of Feynman diagrams from a generating functional through the Freeman-Dyson differential equations, is (almost manifestly) corecursive (\cite{Klafun}).

\subsection{Necessary optimisation of the multiplication}

\noindent
Despite the visual simplicity of the form (\ref{trivmul}), the complexity of the multiplication (and thus, of several other procedures) is disastrous. The computation time rises exponentially with the derivative order, since we have here a typical example of the ``Fibonacci syndrom'': repeated evaluation of identical expressions.
The number of additive terms doubles with each multiplication, and so, with the differentiation order. If one needs just 10 derivatives, there are no serious problems with the schema, the execution is relatively fast, and the memory usage -- acceptable, unless the original expression is convoluted, containg multiple, iterated multiplications and/or divisions. However, if one demands, say, the 200-th derivative, the package is -- in general -- not suitable even for simple sources which do not generate the ugly numeric overflows or NaNs.  We know that 
\begin{equation}
(x\cdot y)^{(n)} = \sum_{k=0}^n {\binom{n}{k} x^{(k)} y^{(n-k)}}\,. \label{dnewton}
\end{equation}
This recipe contains indices and finite sums, and its conversion into the functional corecursive style needs some work. First, we precompute the binomial coefficients, the infinite Pascal triangle as a list of lists, which is available on Internet in several versions, routinely longer than this one:
\vspace{-0.25cm}
\begin{Verbatim}[fontfamily=courier,fontseries=b]
binoms = [1]:map (\b->zipWith (+) (0:b) (b++[0])) binoms
\end{Verbatim}
\vspace{-0.25cm}
Zipping two or more sequences is easy, but the convolution needs two antithetic streams, and since the chains $x$ and $y$ are infinite, they are not invertible (globally). We don't want to manipulate the indexing, analogous to {\tt(!\!!)} either. But we can invert one of the streams incrementally, as is usually done with lists. The convolution proceeds chunk by chunk, $n$ is always the length of the current row of {\tt binoms}, which limits the length of the current zip, and for each increasing $k$, one element from the head of $y$ is transported to the head of an auxiliary list, which thus contains the needed elements of $y$ in reverse order. The inefficient multiplication 1-liner should be replaced by
\vspace{-0.25cm}
\begin{Verbatim}[fontfamily=courier,fontseries=b,commandchars=\\\{\}]
xs * ys = convloop binoms xs ys zeros where
  convloop [] _ _ _ = zeros
  convloop (b:bq) x (y0:>yq) aux =
  let an = y0:>aux
  in dzip3 b x an :> convloop bq x yq an
\end{Verbatim}
\vspace{-0.25cm}
We optimized also the division, simply reconstructed from the inversion of (\ref{dnewton}), and using a trimmed {\tt binoms}:
\vspace{-0.25cm}
\begin{Verbatim}[fontfamily=courier,fontseries=b,commandchars=\\\{\}]
\itshape\color{gry}-- Skipped: first row, & last diagonal of binoms
bint = map init (tail binoms) 

x@(x0:>xq) / y@(y0:>yq) = w where
  w = (x0/y0) :> divloop bint xq yq zeros
  divloop (b:binq) p@(p1:>pr) y@(y1:>yr) t = 
    let yt = (y1:>t) 
    in  (p1 - dzip3 b yt w)/y0 :> divloop binq pr yr yt
\end{Verbatim}
\vspace{-0.25cm}
There is a fair amount of space for variants. The division can be automatically inferred from {\tt recip}, and we could add the de l'Hôpital rule for the special treatment of $0/0$, but this doesn’t need to satisfy every user\ldots{} In any case, all the typical precautions concerning the possible overflows or underflows, or forming of NaNs by the primitive operations should be taken seriously. In the ``real world'' the derivative chains of most interesting functions get corrupted relatively fast, and computing hundreds of derivatives is really rare\ldots

The construction of this and similar samples was natural, but not immediate. Nevertheless while the first version of the technique choked on the expression $\exp(-x)\cdot \sin(x)$ near the $25$-th term, with the above optimisation, 1000 elements have been generated fairly fast.

We end this section by one more example in order to signalise a possible trap. Using the differential recurrence (\ref{hermit}), the code for the Hermite function is the plain transcription of the recursive (not corecursive) definition
\vspace{-0.25cm}
\begin{Verbatim}[fontfamily=courier,fontseries=b,commandchars=\\\{\}]
herm n x = hd (hr n) where  \itshape\color{gry}--The head of the chain
  y = dvar x                \itshape\color{gry}-- y: non-local in hr
  hr 0 = exp(-0.5*>(y*y))
  hr n = (y*z-df z)/(sqrt(2*fromInteger n))\itshape\color{gry}-- df is tail
    where z=hr (n-1)
\end{Verbatim}
\vspace{-0.25cm}\label{hermdef}
In many places the tag {\tt C} in instances of the chain constructors can be omitted, since the automatic type inference/coercion, or the usage of {\tt fromInteger} can help the programmer, but this may be deceiving. 

The execution of {\tt herm 300 x} takes about 20 seconds. But adding just one character speeds-up the execution by a factor of 3.5, to 5.6 seconds. Changing the denominator in the 4-th line above into {\tt C(sqrt(2*fromInteger n))} forces the compiler to {\em understand this expression as a constant}\footnote{With this modification, the manifest coercion {\scour fromInteger} (as used here, not always) becomes redundant.}, precludes the lifting of this number -- the result of the square root -- into the domain of chains, which would imply employing the slower version of division. Perhaps some future Haskell compiler will be able to subsume this property, but this is not simple.

\subsection{Reversion and composition}

\noindent
The differential properties of the composition: $h(x)= g(f(x))$ have been thoroughly analysed, there is a complete formula which gives the derivatives of $h$ through the derivatives of $f$ and $g$. The formula, which will not be quoted here, keeps -- historically -- the name of Faà di Bruno (\cite{Dibruno}), but it has been discovered by Louis Arbogast (\cite{Arbogast}). See also (\cite{Johnson,Flanders}). It contains some complicated combinatorics, and for simple numerical work may be considered too fastidious, but the publications whose authors simplify and reformulate it, continue to appear until today.

\mnd
The reversion of the function $f$ at some given $x$, is another function $g$, such that $g(y)=x$ if $y = f(x)$. The first derivative is a school exercise: 
\begin{equation}
\frac{d}{dx} g(f(x)) = 1 = g'(y)\cdot f'(x)\,.	
\end{equation}
So, $g'(y) = 1/f'(x)$; note that the points of evaluation are different (and we omit them below). Higher derivatives become rapidly more ramified:
\begin{align}
g'' &= -f''/(f')^3\,, \\
g^{(3)} &= \left(3(f'')^2 - f' f^{(3)}\right)/(f')^5\,, \\
g^{(4)} &= \left(-15(f'')^3 + 10 f' f'' f^{(3)} - (f')^2 f^{(4)}\right)/(f')^7\,,\\
g^{(5)} &= \left(105(f'')^4 - 105 f'(f'')^2 f^{(3)} + 10(f')^2 (f^{(3)})^2 + 15 (f')^2 f'' f^{(4)} - \right . \\
& \left .(f')^{(3)} f^{(5)} \right)/(f')^9\,, \quad\text{\ldots{} etc.}
\end{align}
The expression for $g^{(10)}$ occupies a good portion of a printed page, and the numerical coefficients become weighty. Such formulae are also well covered by the literature, published or left as draft, e.g., (\cite{Reynolds}, \cite{Apostol}, or  \cite{Liptaj}).
Since the reversion is simpler and easier (at least here) than the composition, we begin with it. Both here and in the case of composition, it is not rare to find in the references, observations that a predecessor committed a small error, or that a computer algebra system has been used to check the author's results. We shall see that operating with derivative chains, has deprived us of the opportunity to insert many bugs, and obviously the main reason is not the wonderful corecursivity, but working with numerical entities, and delegating the algebra to the code.

What would happen if we replaced $f', g'$, etc. by the complete chains (which will be noted $f_1, g_1, g_2,\ldots$): $g_1 = 1/f_1$, where $f_1 = [f', f'',\ldots]$? Syntactically this is a well-formed construction, chains can be divided, and higher derivatives are assembled without difficulties. Unfortunately its semantics is wrong, because $y \neq x$, and we cannot operate upon two different differentiation variables in one expression, $\ldots g_k(y)\cdots f_m(x)\ldots$ However, we can apply and iterate the differentiation chain rule: 
\begin{equation}
\frac{d}{dy} F(x) = \frac{dx}{dy} \cdot \frac{d F(x)}{dx}\,.
\end{equation}
The Haskell code for the reversion of a function may adapt its form depending on the structure of the source data. For a testing example, we shall take a pair of reciprocally reverse simple functions
\vspace{-0.25cm}
\begin{Verbatim}[fontfamily=courier,fontseries=b,commandchars=\\\{\}]
f x = x/(1+x)\,;   g y = y/(1-y); \quad \itshape\color{gry}\text{-- and } 
x=3\%4; y=f x
\end{Verbatim}
\vspace{-0.25cm}  
So {\tt xs=dvar x} is $\displaystyle{\left[\frac{3}{4},1,0, 0,\ldots\right]}$, and {\tt ys=f xs}: $\displaystyle{\left[\frac{3}{7},\frac{16}{49},\frac{-128}{343}, \frac{1536}{2401},\ldots\right]}$. Having defined
\vspace{-0.25cm}
\begin{Verbatim}[fontfamily=courier,fontseries=b,commandchars=\\\{\}]
revchain f x = x :> revch g1  where
  g1 = 1 / df (f (dvar x))
  revch (h1 :> hq) = h1 :> revch (g1*hq)
\end{Verbatim}
\vspace{-0.25cm}
The call {\tt revchain f x} yields $\displaystyle{\left[\frac{3}{4},\frac{49}{16},\frac{343}{32}, \frac{7203}{128},\ldots\right]}$, identical with {\tt g (dvar y)}. From sin we obtain arcsin, etc. The function {\tt revch} implements the recursive changing of the differentiation variable.
We are far from proud declaring that our three lines of code replace many dozens pages in cited papers. The whole differentiation package behind these lines is involved. We need the compiler environment, and the Haskell runtime support.

We often {\itbf need} analytical, symbolic constructions, in order to reckon the asymptotics, to analyse the singularities or the convergence, to find visible recurrent patterns, and in general -- to gain some insight or find some beauty in  unwieldy formulae. So, our respect for the researchers who process symbolic mathematics is sincere and profound. But we have also seen the results of some white nights of manual work, or of Maple scripts, only to blindly feed them into a plain Fortran code, and we couldn't forget the impression that some precious human time has been wasted.

Applications of the reverse (or inverse) functions are abundant, and they are far from the Internet not very serious remarks, that such object serves to cancel the action of the original operation. The Legendre transforms, which establish the relation between conjugate variables in mechanics (canonical transformations) or in thermodynamics, are ubiquitous, and often they need to be reversed. We might begin with the planning of the trajectory of a robot (or a satellite, etc.), and the inverse kinematics tools set up the recipe for the driving forces and torques, consistent with the planned movement. This is a living subject.

Some historical perspective of the functional reversion can be found in (\cite{Wheeler}). The Reader might verify also that from the definition of the reverse Lambert function, which is elementary: $w\cdot \exp(w) = x$, the program
\vspace{-0.25cm}
\begin{Verbatim}[fontfamily=courier,fontseries=b,commandchars=\\\{\}]
xe x = x*exp(x); ww = revchain xe 0.0
\end{Verbatim}
\vspace{-0.25cm}
builds up the chain identical with {\tt lw0} on page \pageref{lambco}.
\mnd
The composition of functions: $h(x) = g(f(x))$, where $f$ and $g$ are available as derivative towers (or series) is more convoluted. Again, the passage between formal formulae and a working code, may be painful, and the study of the problem continues, since the researches really want (and students really need\ldots) to have something more readable and manageable, than the Faà di Bruno elaborate combinatoric recipe. There are articles containing some chosen information, e.g., the Web collection of formulae offered by Kano Kono (\cite{Alien}).
\begin{align}
h' & = g' f'\,,  \label{hprim}\\ 
h'' & = g' f'' + g''(f')^2\,,\\
h^{(3)} & = g' f^{(3)} + 3 g'' f'' f' + g^{(3)} (f')^3\,, \label{hthird}\\
h^{(4)} & = g' f^{(4)} + g'' \left(4f^{(3)} f' + 3 (f'')^2\right) + 6g^{(3)} f'' (f')^2 + g^{(4)} (f')^4\,,   \label{hp4}
\end{align}
etc. Since, again, $g^{(n)}$ depend on $y$, and the derivatives of $f$ are functions of $x$, we cannot easily mix the chains of $f$ and $g$, and multiply or add them. But all the $f$-dependent coefficients multiplying $g^{(n)}$ {\em can} be processed automatically. We begin with the chain $g_1 f_1$, and we show how to process, say, the segment (\ref{hthird}): $h_3 = g_1 f_3 + 3 g_2 f_2 f_1 + g_3 (f_1)^3$, in order to forge the chain instance of (\ref{hp4}). All lines contain fragments  $\ldots g_k\cdot P_k[f]\ldots$. The recurrence $h_n \to h_{n+1}$ will create here the fragments:  $ g_k P_{k+1}[f] + g_{k+1}\cdot f_1 P_k[f]$. 

The fragments with the same $g_k$ may, and should be immediately combined: $g_k\left(P_{k+1} + f_1 P_{k-1}\right)$, with the exception of the first, and the last $P$ in the chunk. We call these operations {\tt diffg} and {\tt fuse}.
The final result retains only the main values (heads) thereof. The resulting program may not be {\em the} shortest, and most probably not the most efficient, but it should be relatively easy easy to understand.  

Since we ``juggle'' with, and restructure fragments of the final answer, we use both types of sequences, the infinite chains, and finite standard lists, with several specific functions, such as {\tt concatMap} or {\tt sum}
\vspace{-0.25cm}
\begin{Verbatim}[fontfamily=courier,fontseries=b,commandchars=\\\{\}]
compchain gs fs=hd gs:map toscal (iterate (fuse.diffg) [f1])  
 where 
  f1 = df fs; lgd=dToList (df gs) \itshape\color{gry} --Normal lists
  diffg = concatMap (\char92s -> [df s, f1*s])
  fuse seg = head seg : map sum  (chunksOf 2 (tail seg))
  toscalar seg = sum (zipWith (*) (map hd seg) lgd)
\end{Verbatim}
\vspace{-0.25cm}
The procedure {\tt toscalar} computes the scalar products of the $g$s, and $f$ polynomials (reduced to single chains) lists. We used {\tt chunksOf}, a procedure imported from the module {\tt Data.List.Split} of the GHC library, which simply splits a list into a list of lists, grouping the neighbouring two elements. 

The code slows down, but polynomially, and  {\tt (compchain gs fs)} with $f(x) = \sin(x) \exp(-x/2);\  g = \cos$, needs about 80 seconds to generate and print 200 items (less than 6 seconds for 100 elements). Going further is useless, for standard Double floating-point numbers on a 64 bit architecture we usually end with infinities and/or  NaNs. The attempt to compose $\cos$ and $\arccos$ begins well, and finally (after 10 -- 15 terms) explodes, because once the floating point routines fabricate inexact zeros, e.g., $\approx 10^{-15}$, these inexact values propagate and degenerate the computation. In such unstable cases, high derivatives require {very high floating precision}, but it will not prevent the catastrophic evolution of algebraic expressions containing negative exponents, or other ``interesting'' functions, which drive the Universe. 

The execution time dependence on the (simple) functions $f$ and $g$ is  weak. The reason of this deceleration is mainly the laziness of the evaluation protocol, and the accumulation of unevaluated thunks during the recursive journey, not just the recursivity. But since the corecursivity is the essential ingredient of our computational scheme, the strictness analysis and a meaningful improvement would be hard. Some cosmetic advances are possible by a judicious usage of {\tt `seq`}, {\tt \$!}, and other similar tools, but they need a very good knowledge of Haskell and numerous time-consuming tests\footnote{Another lazy language: Clean (\cite{Clean}) has different strictness enhancing tools, and in our opinion, it should be better known by these functional programming community members, who are more interested in concrete applications than in general abstractions, where Haskell is more developed.}.

\section{Power series}

\noindent \label{serpow}
This subject is partly treated in cited papers (\cite{DougP1,DougP2,LPower,Knuth}), and others, but this text is devoted to  several lazy techniques, which covers some more details. For a comprehensive treatise on series, expansions and high derivatives of numerous composite functions in a traditional setting, we recommend the book (\cite{Schwatt}; written in 1924, and reprinted many times).

We include a few comments on the {\em algorithmic} relation between chains and series (see \cite{Carserdif}), which are also lists of coefficients $u_k$ in $U = \sum_{k=0}^\infty{u_k\cdot (x-x_0)^k}$, where $x$ is a formal variable, and $x_0$ will be a constant defined externally (shown in examples). The data structure used by us is similar to derivative towers, with different ``consing'' operator.
\vspace{-0.25cm}
\begin{Verbatim}[fontfamily=courier,fontseries=b]
infixr 5 :-
data Series a = !a :- Series a deriving (Functor,Foldable)
\end{Verbatim}
\vspace{-0.25cm}  
\label{pagdoz}
The choice of arbitrary central point is important if we want to use series as an alternative structure -- the Taylor series -- for derivative chains, with $u_k = d^k/dx^k U(x)\big|_{x_0}$. Nevertheless, in formal manipulations, $x_0$ is almost invisible, and we write simply ``$x$''.
We may note $U = u_0 + x\cdot \ovl{U}$ as {\tt u=(u0:-uq)}. As for chains, the first element of a series identifies its constant value for $x=x_0$, and we have the ``variable'', whose importance is (usually) minor: {\tt svar} means [0, 1, 0, 0\ldots].
In fact, the definition of all constants may be simplified: the {\tt Num} instance of series may define: {\tt fromInteger n = fromInteger n :- 0}, and since {\em here} the type of 0 is Series, {\tt fromInteger} is called recursively, and induces the creation of the infinite tower of zeros. In order to avoid this process, it is better to replace 0 by {\tt szero}, where {\tt szero = 0 :- szero}. The {\em variable} is defined in the source as {\tt svar = 0 :- 1}, which might be slightly confusing.

The addition is linear (zippingWith {\tt(+)}, called here {\tt szip}), and the recursive algorithm for multiplication $U\cdot V$ becomes:
\vspace{-0.25cm}
\begin{Verbatim}[fontfamily=courier,fontseries=b]
(u0:-uq)*v@(v0:-vq) = u0*v0 :- u0*-vq + v*uq
\end{Verbatim}
\vspace{-0.25cm}
Similar to {\tt (*>)} for chains, the operator {\tt (*-)} denotes the multiplication by a scalar, {\tt fmap (*)}, if we specify the derivation of the class {\tt Functor} for the datatype {\tt Series}. In opposition to the multiplication of chains (Leibniz rule, (\ref{trivmul})), the complexity here does not explode exponentially. The division may be coded as
\vspace{-0.25cm}
\begin{Verbatim}[fontfamily=courier,fontseries=b]
(u0:-uq) / v@(v0:-vq) = let w0 = u0/v0
                        in  w0:-(uq - w0*-vq)/v 
\end{Verbatim}
\vspace{-0.25cm}
We may need a few auxiliary functionals, typical in the realm of lists:
\vspace{-0.25cm}
\begin{Verbatim}[fontfamily=courier,fontseries=b,commandchars=\\\{\}]
stl (_:-xq)
sZip op (x:-xq) (y:-yq) = op x y :- sZip op xq yq
sToList (x :- xq)=x:sToList xq \itshape\color{gry}--and its inverse sFromList
\end{Verbatim}
\vspace{-0.25cm}
etc. The coding of elementary functions is somewhat more elaborate than for chains, and is presented e.g., in {\em The Art of Computer Programming}. We define the formal differentiation (over $x$) and integration functions
\vspace{-0.25cm}
\begin{Verbatim}[fontfamily=courier,fontseries=b,commandchars=\\\{\}]
nats = nt 1 where nt n = n:-nt (n+1) \itshape\color{gry}-- Natural numbers
sdif (_:-sq)= sZip (*) sq nats
sint c ss = c :- sZip (/) ss nats
\end{Verbatim}
\vspace{-0.25cm}
(The alternative definition of {\color{red}\tt nats = sFromList [1 ..]} is not appropriate, since this constraints the integers (and their combinations) by the {\tt Enum} class, sometimes too rigid.)

There is a worth mentioning feature of the last definition: while the differentiation needs that the argument -- at least the first element -- be defined, the integration needs the known constant, but the stream argument may be latent, so it can be referred into, inside a corecursive definition, since it is accessed after its (partial) instantiation. The examples follow, in the following list, short and incomplete, {\em to be compared with the functions in the domain of chains}. \label{aritser}
\vspace{-0.25cm}
\begin{Verbatim}[fontfamily=courier,fontseries=b,commandchars=\\\{\}]
exp u@(u0:-_) = w where w = sint (exp u0) (sdif u * w)
log u@(u0:-_) = sint (log u0) (sdif u / u)
sqrt u@(u0:-_)= w where w=sint (sqrt u0) ((1/2)*-(sdif u/w))
sin u@(u0:-_) = sint (sin u0) (sdif u * cos u)
atan u@(u0:-_) = sint (atan u0) (sdif u / (1+u*u)) 
\end{Verbatim}
\vspace{-0.25cm}
etc. This is the essential core of this series package.

\subsection{Example, series for the Lambert ``W'' function} \label{lambsect}

\begin{figure}    
\centering
	\includegraphics[width=0.9\textwidth,keepaspectratio]{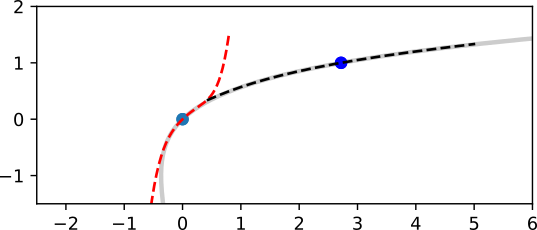}
	\caption{The Lambert W function approximations by series.}
		\label{lambw}
	\vspace{-4mm}
\end{figure}

\noindent
The recurrences for the generation of functions follow patterns similar to those for chains, but used a little differently; the integration is important not only to provide some numerical denominators. We will exercise the  formula (\ref{wladf}). We define the following function, parameterized by the value taken by $W$ for $x-x_0=0$.
\vspace{-0.25cm}  
\begin{Verbatim}[fontfamily=courier,fontseries=b] 
swl w0 = wlx where
    wlx = sint w0 (exp(-wlx)/(1+wlx))
\end{Verbatim}
\vspace{-0.25cm}
From the reverse function, $y = x\cdot \exp(x)$ we see that $y(0)=0$, and $y(1)=e$.These assignments are inverted; the points marked on Fig. \ref{lambw} correspond to the central points of two instances of $W$, where the series have been calculated ({\tt swl 0} and {\tt swl e}) and plotted. 

The figure is just the illustration of the approximation by two Taylor series of 5-th order\footnote{The plot of the curve is exact, but, of course, needn't the Lambert function, its reverse suffices.}. 

\noindent
If truncated numerical series are used to approximate complicated, non-elementary functions, such as shown on Fig. \ref{lambw}, the lazy formulations of the iteration loops with the control of convergence, induce short and readable codes. We shall return to this subject.

\vspace{-3mm}

\subsection{Composition, and reversion of series}

\noindent   
The composition of series is considerably simpler than the case of chains, and is known for many years. From now on we choose $x_0=0$, but here, this is not very important. We present here simple and short algorithms, but there are several important works on optimisation, see e.g., (\cite{Brent})\footnote{But their approach exploit the truncations quite intensely.}.\label{cival}

\mnd
If $U(x) = \sum_{k=0}^n u_k x^k$,  and $V(x) = \sum_{k=1}^n v_k x^k$, 
the series $W=U(V)$ is simply 
$W(x) = \sum_{k=0}^n u_k V(x)^k$, and it is obvious that if $V$ contains $v_0 \neq 0$, $W$ cannot be -- in general -- effectively constructed, since it would require the infinite summation independent of $x$,  $\sum_k {u_k (v_0)^k}$. Without this free term, we rewrite the composition as the infinite  (right-associative) Horner scheme:
\[
U(V)= u_0 + V\cdot(u_1 + V\cdot(u_2 + V\cdot(\cdots)\cdots ))\,.
\]
Since $V^k$ contributes only to the $x^k$ term and higher of $W$, the following code is correctly corecursive. The series {\tt V} can be written as {\tt x*vq}, and we have
\vspace{-0.25cm}
\begin{Verbatim}[fontfamily=courier,fontseries=b,commandchars=\\\{\}]
scompose u (_:-vq) = cmv u where 
  cmv (u0:-uq) = u0 :- vq*cmv uq
\end{Verbatim}
\vspace{-0.25cm}
Notice that the derivative chains don't have the above-mentioned restriction, since the constant ``free term'' is irrelevant.

Chains and series are homologous, and the conversion between these data reduce to correct the factorials:
\vspace{-0.25cm}  
\begin{Verbatim}[fontfamily=courier,fontseries=b]
serToDf s = sdloop s 1 1 where
  sdloop (y:-sq) f n = f*y :> sdloop sq (f*n) (n+1)
\end{Verbatim}
\vspace{-0.25cm}
and similarly (with division) in the opposite direction.

\mnd
The reversion of series is also restricted. If we look for $W$ such that $W(V)=1$ for any given $V$, with some constant term, the problem may be numerically intractable; we will consider $V$ without the first term. Since the value of $v_1$ is of no importance (if not zero), since we can divide $W$ by this coefficient, and correct the answer later, we will solve for $t$ the  equation
\begin{equation}
z = t + U(t)\,,  \label{revequ}
\end{equation}
with $U(t) = u_2 t^2 + u_3 t^3 +  \cdots$ We write thus:
\begin{equation}
t = z - t^2 V(t)\,,    \label{equinv}
\end{equation}
where $V(t)$ is a full series, beginning with the constant term $v_0=u_2$.  This is a common (and important!) case in many scientific and engineering applications, where we have to manipulate the Legendre  transformations (see e.g., \cite{Legendr}).

Our aims are limited, we want just to exercise the laziness, a bit obove the level of the ``primary school''\ldots{}  We may be tempted to interpret the equation (\ref{equinv}) as an algorithm, and hope that the Haskell runtime will be able to reconstruct incrementally the composition $V(t(z))$. Now, the variable $t$ is a series, and $z$ is {\em the} variable ({\tt svar}). The attempt to code: {\color{red}\tt t = z - t*t*scompose v t} fails, we are hit by the bottom, this formula is not correctly corecursive\footnote{If the Reader doesn't see it, compare {{\scour t = (0:-1) - t*t}} with {{\scour t = 0:-p where p = 1 - t*t}}. Check the head value of {\scour t} in both cases.}.

The Reader should understand that there is a significant difference between $a + z\cdot b$, where $z$ is the variable, and $a :\!- b$, {\itbf if} $b$ is an expression which recursively refers to \ldots, well, anything, since the arithmetic in Haskell (as almost everywhere else), is strict. Handling simultaneously strict operations and lazy structuring, needs some care. The solution introduces $w$ such that $t = w\cdot z$, and:
\vspace{-0.25cm}
\begin{Verbatim}[fontfamily=courier,fontseries=b,commandchars=\\\{\}]
sreverse (_ :- _ :- v) = t where \itshape\color{gry} -- u0=0; u1=1; then v
  t = 0 :- w    \itshape\color{gry}                  -- t has no free term
  w = 1 :- (-w*w*scompose v t)
\end{Verbatim}
\vspace{-0.25cm}
For simple functions, such as $f(x)=x/(1+x)$, or $\sin(x)$, series of 100 elements come almost instantaneously, but if the answer is just one point, and we need many $x$, e.g., a full trajectory, we should not neglect the complexity issues, see cited (\cite{Brent}), and also (\cite{Johanss}, who elaborates a fast algorithm based on the Lagrange inversion formula).  For pedagogical purposes, we present the lazy coding of a ``fast'' (allegedly) algorithm, which is nice, but very slow. 

\subsection{Slow reversion by Newton's method}

\noindent
The aim of this non-exercise is a warning, signalled already in the introduction. Given the equation $g(t)=0$, and a sufficiently good initial value, $t_0$, the iteration $t_{n+1} = t_n - g(t_n)/g'(t_n)$ converges quadratically. Here, the equation variables are series, and the equation will take the form $f(t) - z = 0$, where the series $f$ includes the linear term.
The initial point $t_0=z$ yields the series with  {\bf one} exact term. The first approximation gives two more correct coefficients, and then four more, eight, etc. The code which might have to be augmented by the {\em choice} of the desired precision, iterates the standard Newton correction:
\vspace{-0.25cm}
\begin{Verbatim}[fontfamily=courier,fontseries=b,commandchars=\\\{\}]
newtreverse f = iterate nxt svar where
  f' = sdif f
  nxt t = t - (scompose f t - svar) / (scompose f' t)
\end{Verbatim}
\vspace{-0.25cm}
Taking {\tt (newtreverse f)!!7} returns a series with 256 exact (machine precision) terms. This is short and readable, but the process may be 10-fold slower than {\tt sreverse} (from 3.5 to almost 40 seconds), and this is easy to understand: the algebraic operations, subtractions, composition and division, are {\em not atomic}, and they become more and more costly, when the consumer establishes the value of the next term, after having repeated the construction of all previous ones. Unevaluated thunks may consume an inordinate amount of memory, and evaluating them later is costly. So, while we sincerely advocate the lazy algorithms, some experience, and a conscious coding discipline is really necessary.

Practically all work on the reversion of series employed some truncation procedures, and Newton himself started with them; the algorithm described in  (\cite{Dunham}) converges linearly. 

\subsection{Asymptotics of factorials, and lazy Stirling approximation}
	
\noindent
This is an exercise inspired by the problem: {\em Perturbation of Stirling Formula} in the book ``Concrete Mathematics'', (\cite{GDekPa}). As elsewhere in this text, our aim is to transform an equation into an iterative algorithm, which can progress  without the presence of truncations. The typical readers of this text are acquainted with the basic Stirling approximation of a factorial $(n!)$ for big values of $n$:
\begin{equation}
n! \approx \sqrt{2\pi n} \left(\frac{n}{e}\right)^n \cdot S\left(\frac{1}{n}\right)\,,
\label{stirfo}
\end{equation}
where $S(1/n) = 1 + s_1/n + s_2/n^2 + \cdots$ is an asymptotic, divergent correction series, whose usage is mainly illustrative, so finding in the popular literature the formulae with more than three $s_m$  terms is not straightforward. We shall need only (\ref{stirfo}), and the basic recurrence of factorials, in order to generate a set of equations for the coefficients $s_m$. We begin with
\begin{equation}
\frac{n!}{(n-1)!} = n = \frac{n}{e}\sqrt{\frac{n}{n-1}}
   \left(\frac{n}{n-1} \right)^{n-1} \frac{S(1/n)}{S(1/(n-1))}\,.
    \label{recstr}
\end{equation}
Introducing a variable $x = 1/n$, in order to work with the series about zero, we rewrite (\ref{recstr}) into
\begin{equation}
S\left(\frac{x}{1-x}\right) = S(x)\cdot G(x)\,, \quad\text{where}\quad
  G(x) = \exp\left(-1 + (1/2 -1/x)\cdot \log(1-x) \right)\,.
\label{recseq}
\end{equation}
In the coding below, $x$ will mean always {\tt svar, (0 :- 1)} completed by the infinite chain of zeros, as on the page (\pageref{pagdoz}), and will be typed as real: floating or rational, which will be commented upon. 
The expression $G$ is singular at zero, but since its limit exists\footnote{We omit the proofs, and we highly recommend the book (\cite{GDekPa}), which will help the reader to understand the used  manipulations; our contribution is mainly the casting of the relevant equations into a corecursive form, and solving them.}, we dismiss $1/x$ in $G$ together with the 0-th term of the logarithm series, which is zero. Concretely, we code {\itbf almost}
\vspace{-0.25cm}
\begin{Verbatim}[fontfamily=courier,fontseries=b,commandchars=\\\{\}]
g = exp(-1 + lo/2 - stl lo) where lo = log(1-x)
\end{Verbatim}
\vspace{-0.25cm}
Why almost? Since fractions are usually considered nicer than floats for human eyes, this is the accepted presentation of many algebraic formulae, but the logarithm and exponential are members of the {\tt Floating} class, and dislike rationals. But the algorithms for these functions (all on the page \pageref{aritser}) are {\em almost} neutral wrt. the real number domain, only the zeroth-term needs the standard floating exp, and log. We constructed thus a few restricted functions, named: {\tt exp0}, {\tt log1}, and {\tt sqrt1} with the digit indicating the value of the zeroth term, which replaces the expression after {\tt sint} in the definitions of {\tt exp} etc. It is easy to check that the linear term of $G$ vanishes, and the result is
\begin{equation}
G(x) = 1 + x^2 F(x)\,, \ \text{with}\  F(x) =  \frac{1}{12} + \frac{1}{12} x + \frac{113}{1440} x^2 + \frac{53}{720} x^3 + \frac{25163}{362880} x^4 + \cdots
\end{equation}
However, the recursive equation of $S$ in (\ref{recseq}) is still not a proper lazy algorithm, the linear term ($s_1$) cannot be computed directly, the equation (\ref{recseq}) yields the bottom $s_1 = s_1$. But, although not immediately visible, this trivial issue can be eliminated by subtracting $S(x)$ from both sides of (\ref{recseq}). We get
\begin{equation}
x s_1\left(\frac{1}{1-x} - 1\right)  + \cdots + x^m s_m\left(\frac{1}{(1-x)^m} - 1\right) + \cdots
= x^2 F(x) S(x)\,.  \label{aftsub}
\end{equation}
All differences in parentheses have one factor $x$ extractible, e.g., $1/(1-x) - 1 = x/(1-x)$, and the factor $x^2$ can be simplified from both sides. We introduce 
\begin{equation}
R_m(x) = (1/(1-x)^m -1)/x\,,
\end{equation} 
which obeys the recurrence $R_{m+1} = (R_m + 1)/(1-x)$. 
The formula (\ref{aftsub}) turns into
\begin{equation}
s_1 R_1 +x s_2 R_3 + x^2 s_3 R_3 + \cdots = F(x) S(x)
\label{simpsa}
\end{equation}
The lowest (free from $x$) term of (\ref{simpsa}), the head of $s_1 R_1$ is equal to the head of $F(x)$, since $s_0=1$. This gives directly $s_1 = 1/12$. But $s_1 R_1$ contains higher orders in $x$, so the next stage of the iteration begins with its transfer into the RHS:
\begin{equation}
s_2 R_2 + x s_3 R_3 + x^2 s_4 R_4 + \cdots = \frac{1}{x} (F(x) S(x) - s_1 R_1(x))\,,
\end{equation}
permitting to compute $s_2$ when $s_1$ is solved. The equation becomes a back-substitution algorithm for $s_m$.
\vspace{-0.3cm}
\begin{Verbatim}[fontfamily=courier,fontseries=b,commandchars=\\\{\}]
stirling = 1 :- backsub (1/(1-x)) (f*stirling) where
  backsub rm rhs = 
    let sm = shd rhs / shd rm
    in  sm :- backsub ((1+rm)/(1-x)) (stl (rhs - sm*-rm))
\end{Verbatim}
\vspace{-0.3cm}
The result is
\begin{equation}
\begin{aligned}
S = &1 + \frac{1}{12} x + \frac{1}{288} x^2 + \frac{-139}{51840}x^3 + \frac{-571}{2488320}x^4 + \frac{163879}{209018880}x^5 + \frac{5246819}{75246796800}x^6 +\\ 
&\frac{-534703531}{902961561600}x^7 + \frac{-4483131259}{86684309913600} x^8 + \cdots\,. \label{labesti}
\end{aligned}
\end{equation}
Our code is quite compact, but the algorithmisation process is not automatic, requires some thinking, and stands on the shoulders of the series package. 

The next example demonstrates another, very classical and compact way to derive this infinite series, but which also needs some head-scratching.

\section{The integration of Laplace, and Stirling approximation}
\noindent
In general case, if we want to compute many terms of the asymptotic evaluation of
\begin{equation}
	I(x) = \int{f(t) e^{-x \varphi(t)} dt}\,,
\end{equation}
for $x \to \infty$, knowing that $\varphi(t)$ has one maximum inside the integration interval, see (\cite{Bender}), or some other comprehensive book on mathematical methods for physicists; see also (\cite{Strawder}). The Laplace method and its variants (saddle point, steepest descent)
are extremely important in natural and technical sciences, in particular in statistical physics. In several cases, e.g. in nuclear physics, the large parameter often corresponds to the number of particles involved, and is finite. So, higher terms of the expansion may be numerically useful.

The technique consists in expanding $\varphi$ about the position of this maximum $t=p$: $\varphi'(p) = 0$. Then $\varphi(t) = \varphi(p) + \varphi''(p)\cdot (t-p)^2/2 + R(t)$. The terms in the exponent above Gaussian: $R(t) = x \varphi'''/6 \cdot (t-p)^3 + \cdots$ will be approximated by a polynomial in $t-p$, and higher powers yield higher negative powers of $x$.

\setlength{\columnsep}{0.3cm}
\begin{wrapfigure}[15]{r}{3.85cm} 
	\centering
	\vspace{-0.6cm}
	\includegraphics[width=3.7cm]{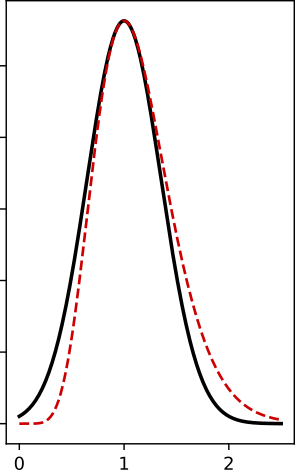}
	\vspace{-0.2cm}
	\caption{Factorial integrand}
	\label{stirlap}
\end{wrapfigure}
The starting point is the integral definition of the factorial, which will be treated by the Laplace technique:
\begin{equation}
n! =\int_0^\infty t^n \exp(-t) dt = 
  \int_0^\infty \exp(n \log t - t) dt\,.
\end{equation}
We shall re-derive the result (\ref{labesti}) above, but here the issue is not the coercion of an equation into a corecursive algorithm, but the handling of an intricate, awkward structure: a series of series, avoiding the usage of symbolic indeterminates. Our goal is a series in $1/n$, but we have to expand the exponent in $t$, before its elimination through the integration.

On the Fig. \ref{stirlap}, the solid line is a Gaussian (giving the zeroth Stirling approximation of the factorial), and the dashed curve is the full integrand for $n=8$: a product of an increasing power, and the decreasing exponential. The classical Laplace technique consists in a development of this integrand around the maximum, which is trivial to calculate, $t = n$, or $z = 1$, where $z$ is an auxiliary variable: $z = t/n$. For big $n$, almost all area under the curve is restricted to the vicinity of $z=1$, and taking the integration along the full real axis is not a serious sin. From the expansion of the exponential argument, we obtain
\begin{equation}
\displaystyle{n^{n+1} \int \exp\left( n (\log z - z) \right) dz =
\frac{n^{n+1}}{ e^{n}} \int \exp\left(-\frac{n z^2}{2}\right)
\exp\left( n z^3\sum_{j=0}^\infty \frac{(-1)^{j} z^j}{j+3} \right) dz}\,.
\label{gauin}
\end{equation}
The code
\vspace{-0.5cm}
\begin{Verbatim}[fontfamily=courier,fontseries=b,commandchars=\\\{\}]
  z = svar + 1 :: Series Rational
  w = stl $ stl $ stl (log1 z - z) -- {\it begins with \(z^3\)}
\end{Verbatim}
\vspace{-0.4cm}
yields for the coefficients sequence (with no explicit $n$, nor $z$) {\tt w}:  {\tt 1 \% 3 :- (-1) \% 4 :- 1 \% 5 :- (-1) \% 6 :-  \ldots]}. Now, not $z$, but $w$ including (implicitly) $n$ will be treated as the series variable, and the result of {\tt en = stl (exp0 (0 :- w :- 0))} is (converted into a list of lists) from a {\tt Series (Series Rational)}:.
\vspace{-0.4cm}
\begin{Verbatim}[fontfamily=courier,fontseries=b,commandchars=\\\{\}]
  [[1 \% 3,(-1) \% 4,1 \% 5,(-1) \% 6,1 \% 7,(-1) \% 8\ldots],
   [1 \% 18,(-1) \% 12,47 \% 480,(-19) \% 180,153 \% 1400,\ldots],
   [1 \% 162,(-1) \% 72,31 \% 1440,(-493) \% 17280,\ldots]\ldots ]
\end{Verbatim}
\vspace{-0.4cm}
The zeroth row of the exponential (the zeroth approximation of the result) equal to 1 is not used and absent; assume that the index of the beginning row is 1.) We will disregard the constant term of the integrand and the Gaussian normaisation, no $\sqrt{2\pi n}\cdot (n/e)^n$ below, the result begins with $1/12$.

This double series is purely numerical, the role of powers of $n$ (and its associated coefficients in $w$) is played by the index of the row, let's say: $p$. The power  of $z$ inside $w$  -- by the column of the array: $j$. In order to reduce these variables and construct a sequence indexed by the power of $1/n$, we need to perform the Gaussian integrations:
\begin{equation}
\int z^{2 m} \exp\left(\frac{-n z^2}{2}\right) = \sqrt{\frac{2 \pi}{n}} \cdot \frac{(2m-1)!!}{n^m}\,.
\end{equation}
This is just a mapping, no need to ``really'' integrate anything. The final step is the sorting and summing  all terms which contribute to the same power of $(1/n)$. It is easy to see that the terms which go together, lie alongs the diagonals $p+j=\text{const}$ (even). The {\em symbolic structure} of the array {\tt en} is plotted on Fig. \ref{sstru}, the values therein  are the powers of $1/n$ to which an element contributes, and the unused elements are empty.

More concretely, the $(p,j)$-th element (with coefficient  $C_{p j}$) of the double series array which undergoes the Gaussian integration as in (\ref{gauin}), is mapped as below:
\begin{equation}
	\int \cdots  [\cdots n^p z^{3p} [\cdots C_{p j} z^j \cdots] \cdots]
	\longrightarrow C_{pj} \cdot (3p+j-1)!! \cdot n^{\displaystyle{-(p+j)/2}}\,.
\end{equation}

\setlength{\columnsep}{0.2cm}
\begin{wrapfigure}[10]{l}{7.0cm} 
	\centering
	\vspace{-0.9cm}
\[
 \begin{array}{|llllllllllll|} 
 	\hline
 \ & 1 &\ & 2 &\ & 3 & \ & 4 & \ & 5 & \ &\cdots\\
 1 &\ & 2 & \ & 3 & \ & 4 & \ & 5 & \ & 6 &\cdots\\	
  \ & 2 &\ & 3 & \ & 4 & \ & 5 & \ & 6 & \ &\cdots\\
 2 & \ & 3 & \ & 4 & \ & 5 & \ & 6 & \ & 7 & \cdots\\
   \ & 3 &\ & 4 & \ & 5 & \ & 6 & \ & 7 & \ &\cdots\\
 3 & \ & 4 & \ & 5 & \ & 6 & \ & 7 & \ & 8 &\cdots\\
 \ & 4 & \ & 5 & \ & 6 & \ & 7 & \ & 8 &\ &\cdots\\
  4 &\ & \multicolumn{9}{c}{\cdots}&\\ 
 \hline
 \end{array}	
\]
	\vspace{-0.2cm}
	\caption{Stirling array: powers of $1/n$.}
	\label{sstru}
\end{wrapfigure}
\noindent
In order to obtain the answer (\ref{labesti}), all entries should be multiplied by the appropriate semi-factorial coefficient, and each diagonal sequence accumulated separately. In principle, for a given $n^{-k}$ with $k=(p+j)/2$, we could simply compute $\sum_{j=0}^{(k-1)} C_{(2k-j)(j)}\cdot (6k-2j-1)!!$. But for lists, the indexing operator has linear complexity, and even if there is no need for many terms, we propose -- for pedagogical reasons -- a more list-oriented approach. We will iterate left shifts of some initial rows followed by the separation of the initial fragments of the first column: symbolically, first ``1 1'', then ``2 2 2 2'' etc.

The final stage of the algorithm uses the list representation trivially reconstructed from the exponential {\tt en}, since the specific series structure, nor its arithmetic are not needed, and Haskell lists are better optimized.

\vspace{-0.4cm}
\begin{Verbatim}[fontfamily=courier,fontseries=b,commandchars=\\\{\}]
  enl = sToList (fmap sToList en)
  dbfacs = 0 : dbf 1 2 [3 ..] where \itshape\color{gry}-- Semi-factorials
    dbf x y (a : b : r) = x : y : dbf (x*a) (y*b) r
  
  tabl = crow 1 enl where 
    crow p ll = ccol p 0 (head ll) : crow (p+1) (tail ll)
    ccol p j l = (dbfacs!!(3*p+j-1))*head l 
               : ccol p (j+1) (tail l)  
\end{Verbatim}
\vspace{-0.4cm}
Here comes the final stage, iterated vertical alignment of diagonal segments.
\vspace{-0.4cm}
\begin{Verbatim}[fontfamily=courier,fontseries=b,commandchars=\\\{\}]
  ss m ll=(f,(map tail f)++s) where (f,s)=splitAt m ll\itshape\color{gry} --Aux.
  shift m ll = snd (ss m ll)
  separ m ll = let (f,b) = ss m ll in (map head f,b)
  
  stirling = diag 1 tabl  where
    diag m tbl = let (d,nxt) = separ (m+1) (shift m tbl)
                 in sum d : diag (m+2) nxt
\end{Verbatim}
\vspace{-0.4cm}
The value of {\tt stirling} is the infinite list of coefficients of the series (\ref{labesti}) , beginning with $1/12$.

\section{Concluding remarks}

\noindent
The content of this paper is related to applied mathematics, but it is not a mathematical paper. The key word in the title is ``coding''.

Actually, it is not just coding in the sense of transcribing the potentially executable material from paper to the input files, but the organisation of those execution instructions, the {\bf algorithmisation} of the task to be solved. 
Lazy functional algorithms deserve to be treated by the community of natural sciences as a useful {\itbf methodology}, not just as ``academic'' tricks, but this requires well designed teaching practices, which evolve slowly. This is why we have chosen the orthodox numerical dimension, although some examples bear a distinct symbolic flavour. 

Our inspiration came from various sources, e.g., from the observation that the popular presentation of scientific algorithms often takes the form of imperative pseudo-code, which we have learnt during the 70-ties of the last century. This is not bad {\em per se}, since the pedagogy of programming should be based on patterns mastered by the instructors, but it means that the standard teaching style evolves rather slowly, and sometimes even retrogresses\footnote{With the advent of new commercialised AI tools, the teaching of computational logic (and Prolog), seems to have diminished in several education establishments.}. Convincing astronomers or biologists to ``think functionally'' is not easy.

\mnd
Some people may remember the comment by Leibniz,  \cite{Leibniz} addressed to astronomers: {\em It is unworthy of excellent men to lose hours like slaves in the labour of calculation which could be relegated to anyone else if machines were used}. However, most of us would probably prefer to cite Agent Smith from the movie Matrix. Plainly: {\em Never send a human to do a machine's job}\ldots

This may be right and wrong, since often our goal is not the final number, e.g., 42, but the understanding of its origins and its distillation. But what might be arguable, is the impression that formulae with symbolic indeterminates (and their mechanical processing) give {\em by definition}, more understanding than the numerical procedures, unfortunately, the human attention may be dispersed on syntactic issues\ldots{}  But maginative human thinking will always remain the essential ingredient of high-level computer programming because we want {\em interesting} approaches to interesting problems.

We believe that the lazy algorithmisation techniques may provide some high-level, and interesting layers of reasoning and understanding, especially when those numbers are assembled into intricate, unwieldy structures.

\bibliography{highder}

\begin{thebibliography}{41}

\bibitem[{Apostol}(2000)]{Apostol}
{Apostol, T.~M.} (2000) Calculating higher derivatives of inverses. \textit{The
  Amer. Math. Monthly}. {\bf 107}(8), 738 -- 741.

\bibitem[{Arbogast}(1800)]{Arbogast}
{Arbogast, L.} (1800) {\em Du calcul des dérivations\/}. Levrault,Strasbourg.
  \url{https://books.google.fr/books?id=YoPq8uCy5Y8C&printsec=frontcover&hl=fr&source=gbs_ge_summary_r&cad=0#v=onepage&q&f=false}.

\bibitem[{Bender and Orszag}(1978){Bender \& Orszag}]{Bender}
{Bender, C. \& Orszag, S.} (1978) {\em Advanced Mathematical Methods for
  Scientists and Engineers\/}. McGraw-Hill.

\bibitem[{Berz}(1991)]{MBerz}
{Berz, M.} (1991) Algorithms for higher order automatic differentiation in many
  variables with applications to beam physics. Breckenridge workshop on
  automatic differentiation. Siam 1991.

\bibitem[{Brent and Kung}(1978){Brent \& Kung}]{Brent}
{Brent, R.~P. \& Kung, H.-T.} (1978) Fast algorithms for manipulating formal
  power series. \textit{Journal of the ACM}. {\bf 25}(4), 581 -- 595.
  \url{http://www.eecs.harvard.edu/~htk/publication/1978-jacm-brent-kung.pdf}.

\bibitem[{Brito et~al.}(2008){Brito {\em et~al.\/}}]{Brito}
{Brito, P. {\em et~al.\/}}. (2008) {Euler, Lambert, and the Lambert W-function
  today}. \textit{The Mathematical Scientist}. {\bf 33}, 203 --– 219.
  \url{https://www.researchgate.net/publication/266167744_Euler_Lambert_and_the_Lambert_W-function_today}.

\bibitem[{Bücker et~al.}(2024){Bücker {\em et~al.\/}}]{Autodiff}
{Bücker, M. {\em et~al.\/}}. (2024) autodiff.org.
  \url{https://www.autodiff.org/}.

\bibitem[{Carothers et~al.}(2012){Carothers {\em et~al.\/}}]{Carserdif}
{Carothers, D.~C. {\em et~al.\/}}. (2012) Connections between power series
  methods and automatic differentiation. In {\em Lecture Notes in Computational
  Science and Engineering\/}. Springer. chapter~1, pp. 1 –-- 10.

\bibitem[{Corless et~al.}(1996){Corless, Knuth, {\em et~al.\/}}]{LambertW}
{Corless, R.~M., Knuth, D.~E. {\em et~al.\/}}. (1996) {On the Lambert W
  function}. \textit{Advances in Computational Mathematics}. {\bf 5}, 329 –--
  359. \url{https://cs.uwaterloo.ca/research/tr/1993/03/W.pdf}.

\bibitem[{Dunham}(2008)]{Dunham}
{Dunham, W.} (2008) {\em The Calculus Gallery: Masterpieces from Newton to
  Lebesgue\/}. Princeton University Press.

\bibitem[{Eager et~al.}(2016){Eager, Pendrill, \& Reistad}]{Eager}
{Eager, D., Pendrill, A.-M. \& Reistad, N.} (2016) Beyond velocity and
  acceleration: jerk, snap and higher derivatives. \textit{European Journal of
  Physics}. {\bf 37}, 1 -- 11.
  \url{https://iopscience.iop.org/article/10.1088/0143-0807/37/6/065008/pdf}.

\bibitem[{{Faà di Bruno}}(1855)]{Dibruno}
{{Faà di Bruno}, F.} (1855) Sullo sviluppo delle funzioni. \textit{Annali di
  Scienze Matematiche e Fisiche}. {\bf 6}, 479 -- 480.
  \url{https://books.google.fr/books?id=ddE3AAAAMAAJ&pg=PA479&redir_esc=y#v=onepage&q&f=false}.

\bibitem[{Flanders}(2001)]{Flanders}
{Flanders, H.} (2001) {From Ford to Faà}. \textit{American Mathematical
  Monthly}. {\bf 108}(6), 558 –-- 561.

\bibitem[{Frabetti and Manchon}(2011){Frabetti \& Manchon}]{FaaDiBruno}
{Frabetti, A. \& Manchon, D.} (2011) {Five interpretations of Faà di Bruno’s
  formula}. {Dyson--Schwinger Equations and Faà di Bruno Hopf Algebras in
  Physics and Combinatorics}. \url{https://arxiv.org/pdf/1402.5551}.

\bibitem[{Graham et~al.}(1994){Graham, Knuth, \& Patashnik}]{GDekPa}
{Graham, R.~L., Knuth, D.~E. \& Patashnik, O.} (1994) {\em Concrete
  Mathematics\/}. Addison-Wesley, Reading.

\bibitem[{Johansson}(2015)]{Johanss}
{Johansson, F.} (2015) A fast algorithm for reversion of power series.
  \textit{Mathematics of Computation}. {\bf 84}, 475 -- 484.
  \url{https://arxiv.org/abs/1108.4772}.

\bibitem[{Johnson}(2002)]{Johnson}
{Johnson, W.~P.} (2002) {The Curious History of Faà di Bruno's Formula}.
  \textit{The American Mathematical Monthly}. {\bf 109}(3), 217 -- 234.
  \url{https://maa.org/sites/default/files/pdf/upload_library/22/Ford/Johnson217-234.pdf}.

\bibitem[{Kaplanski}(1957)]{Kapla}
{Kaplanski, I.} (1957) {\em An Introduction to Differential Algebra\/}.
  Hermann, Paris.
  \url{http://mmrc.iss.ac.cn/~weili/DifferentialAlgebra/References/Kaplansky.pdf}.

\bibitem[{Karczmarczuk}(1994)]{Klafun}
{Karczmarczuk, J.} (1994) Lazy functional programming and manipulation of
  perturbational series. {New Computing Techniques in Physics Research III}.
  {World Scientific}.

\bibitem[{Karczmarczuk}(1997)]{LPower}
{Karczmarczuk, J.} (1997) Generating power of lazy semantics.
  \textit{Theoretical Computer Science}. {\bf 187}, 203 --– 219.

\bibitem[{Karczmarczuk}(1998)]{Icfp}
{Karczmarczuk, J.} (1998) Differentiation of functional programs. Proc. III ACM
  Intern. Conference on Functional Programming. ACM. pp. 195 -- 203.

\bibitem[{Karczmarczuk}(2001)]{Hosc}
{Karczmarczuk, J.} (2001) Functional differentiation of computer programs.
  \textit{Higher-Order and Symbolic Computation}. {\bf 14}, 35 -- 57.

\bibitem[{Knuth}(1997)]{Knuth}
{Knuth, D.~E.} (1997) {\em The Art of Computer Programming, Seminumerical
  Algorithms\/}. Addison-Wesley.

\bibitem[{Kono}(2017)]{Alien}
{Kono, K.} (2017) Higher derivative of composition.
  \url{https://fractional-calculus.com/higher_derivative_composition.pdf}.

\bibitem[{Leibniz~(von)}(1685)]{Leibniz}
{Leibniz~(von), G.~W.} (1685) Comment on the {Step Reckoner}, his calculating
  machine invented in 1673.

\bibitem[{Liptaj}(2017)]{Liptaj}
{Liptaj, A.} (2017) Higher order derivatives of the inverse function.
  \url{https://vixra.org/pdf/1703.0295v1.pdf}.

\bibitem[{Martins et~al.}(2003){Martins, Sturdza, \& Alonso}]{CsMartins}
{Martins, J.~J., Sturdza, P. \& Alonso, J.~J.} (2003) The complex-step
  derivative approximation. \textit{ACM Trans. on Math. Soft.} {\bf 29}, 245 --
  262. \url{https://hal.science/hal-01483287/document}.

\bibitem[{McIlroy}(1990)]{DougP1}
{McIlroy, M.~D.} (1990) Squinting at power series. \textit{Software: practice
  and Experience}. {\bf 209}, 661 -- 683.

\bibitem[{McIlroy}(1999)]{DougP2}
{McIlroy, M.~D.} (1999) Power series, power serious. \textit{Journal of
  Functional Programming}. {\bf 9}, 323 -- 335.

\bibitem[{Mező}(2024)]{Mezo}
{Mező, I.} (2024) {References on the Lambert W function and its
  generalizations}.
  \url{https://sites.google.com/site/istvanmezo81/references-on-the-lambert-w-function}.

\bibitem[{Plasmeijer et~al.}(2022){Plasmeijer {\em et~al.\/}}]{Clean}
{Plasmeijer, R. {\em et~al.\/}}. (2022) Clean.
  \url{https://wiki.clean.cs.ru.nl/Clean}.

\bibitem[{Reynolds}(1944)]{Reynolds}
{Reynolds, J.~B.} (1944) Reversion of series with application. \textit{The
  Amer. Math. Monthly}. {\bf 51}, 578 -- 580.

\bibitem[{Ritt}(1966)]{Ritt}
{Ritt, J.~F.} (1966) {\em Differential Algebra\/}. Dover, New York.
  \url{http://mmrc.iss.ac.cn/~weili/DifferentialAlgebra/References/Ritt.pdf}.

\bibitem[{Schwatt}(1962)]{Schwatt}
{Schwatt, I.~J.} (1962) {\em An Introduction to the Operations with Series\/}.
  Chelsea scientific books. Chelsea Publishing Company.
  \url{https://archive.org/details/anintroductionto0000ijsc/page/n7/mode/2up}.

\bibitem[{Squire and Trapp}(1998){Squire \& Trapp}]{Squitra}
{Squire, W. \& Trapp, G.~E.} (1998) {Using Complex Variables to Estimate
  Derivatives of Real Functions}. \textit{{SIAM Rev.}} {\bf 40}(1), 110 --–
  112.
  \url{https://researchrepository.wvu.edu/cgi/viewcontent.cgi?article=1425&context=faculty_publications}.

\bibitem[{Strawderman}(2000)]{Strawder}
{Strawderman, R.} (2000) {Higher-Order Asymptotic Approximation: Laplace,
  Saddlepoint, and Related Methods}. \textit{Journal of the American
  Statistical Association}. {\bf 95}(452), 1358 -- 1364.
  \url{https://www.researchgate.net/publication/254287979_Higher-Order_Asymptotic_Approximation_Laplace_Saddlepoint_and_Related_Methods}.

\bibitem[{Visser}(2004)]{Visser}
{Visser, M.} (2004) Jerk, snap and the cosmological equation of state.
  \textit{Classical and Quantum Gravity}. {\bf 21}(11), 2603 -- 2616.
  \url{https://arxiv.org/pdf/gr-qc/0309109.pdf}.

\bibitem[{Wheeler}(2017)]{Wheeler}
{Wheeler, N.} (2017) Functional inversion strategies.
  \url{https://www.reed.edu/physics/faculty/wheeler/documents/Miscellaneous\%20Math/Functional\%20Inversion\%20Strategies/Applied\%20Functional\%20Inversion\%20.pdf}.

\bibitem[{Wikipedia}(2023)]{WickAD}
{Wikipedia}. (2023) Automatic differentiation.
  \url{https://en.wikipedia.org/wiki/Automatic\_differentiation}.

\bibitem[{Wikipedia}(2023)]{Crackle}
{Wikipedia}. (2023) Snap, crackle and pop.
  \url{https://en.wikipedia.org/wiki/Snap,\_Crackle\_and\_Pop}.

\bibitem[{Zia et~al.}(2009){Zia, Redish, \& McKay}]{Legendr}
{Zia, R.~K., Redish, E.~F. \& McKay, S.~R.} (2009) Making sense of the legendre
  transform. \textit{Am. J. Phys.} {\bf 77}, 614 -- 622.
  \url{https://www3.nd.edu/~powers/ame.20231/zia.pdf}.

\end{thebibliography}

\end{document}